\documentclass[aps,pra,preprint,groupedaddress]{revtex4-1}

\usepackage{amsmath}
\usepackage{amssymb}
\usepackage{mathrsfs}
\usepackage{graphicx}
\usepackage{color}
\usepackage{multirow}

\begin{document}


\title{Quantum walks on regular uniform hypergraphs}
\author{Ying Liu, Jiabin Yuan, Bojia Duan, Dan Li}
\affiliation{College of Computer Science and Technology, Nanjing University of Aeronautics and Astronautics, No.29 Jiangjun Avenue, 211106 Nanjing, China.}

\date{\today}

\begin{abstract}

    Quantum walks on graphs have shown prioritized benefits and applications in wide areas. In some scenarios, however, it may be more natural and accurate to mandate high-order relationships for hypergraphs, due to the density of information stored inherently. Therefore, we can explore the potential of quantum walks on hypergraphs. In this paper, by presenting the one-to-one correspondence between regular uniform hypergraphs and bipartite graphs, we construct a model for quantum walks on bipartite graphs of regular uniform hypergraphs with Szegedy's quantum walks, which gives rise to a quadratic speed-up. Furthermore, we deliver spectral properties of the transition matrix, given that the cardinalities of the two disjoint sets are different in the bipartite graph. Our model provides the foundation for building quantum algorithms on the strength of quantum walks on hypergraphs, such as quantum walks search, quantized Google's PageRank, and quantum machine learning.
\end{abstract}
\maketitle

\section{Introduction}
\label{sec1:level1}

     As a quantum-mechanical analogs of classical random walks, quantum walks have become increasingly popular in recent years, and have played a fundamental and important role in quantum computing. Owing to quantum superpositions and interference effects, quantum walks have been effectively used to simulate quantum phenomena \cite{bracken2007free}, and realize universal quantum computation \cite{childs2009universal,lovett2010universal}, as well as develop extensively quantum algorithms \cite{venegas2012quantum}. A wide variety of discrete quantum walk models have been successively proposed. The first quantization model of a classical random walk, which is the coined  discrete-time model and which is performed on a line, was proposed by Aharonov \textit{et al}. \cite{aharonov1993quantum} in the early 1990s. Aharonov later studied its generalization for regular graphs in Ref. \cite{aharonov2001quantum}. Szegedy \cite{szegedy2004quantum} proposed a quantum walks model that quantizes the random walks, and its evolution operator is driven by two reflection operators on a bipartite graph. Moreover, in discrete models, the most-studied topology on which quantum walks are performed and their properties studied are a restricted family of graphs, including line \cite{nayak2000quantum,portugal2015one}, cycle\cite{bednarska2003quantum,melnikov2016quantum},  hypercube\cite{moore2002quantum,potovcek2009optimized}, and general graphs\cite{chakraborty2016spatial,wong2015faster,krovi2016quantum}. Indeed, most of the existing quantum walk algorithms are superior to their classical counterparts at executing certain computational tasks, e.g., element distinctness \cite{ambainis2007quantum,belovs2012learning}, triangle finding \cite{ magniez2007quantum,lee2013improved}, verifying matrix products \cite{buhrman2006quantum}, searching for a marked element \cite{shenvi2003quantum,krovi2016quantum},quantized Google's PageRank\cite{paparo2012google} and graph isomorphism \cite{douglas2008classical,gamble2010two,berry2011two,wang2015graph}.

     In mass scenarios, a graph-based representation is incomplete, since graph edges can only represent pairwise relations between nodes. However, hypergraphs are a natural extension of graphs that allow modeling of higher-order relations in data. Because the mode of representation is even nearer to the human visual grouping system, hypergraphs are more available and effective than graphs for solving many problems in several applications. Owing to Zhou's random walks on hypergraphs for studying spectral clustering and semi-supervised ranking\cite{zhou2007learning}, hypergraphs have made recent headlines in computer vision \cite{ yu2014high,huang2017effect}, information retrieval\cite{hotho2006information,yu2014click,zhu2017unsupervised}, database design \cite{jodlowiec2016semantics}, and categorical data clustering \cite{ochs2012higher}. Many interesting and promising findings were covered in random walks on hypergraphs, and quantum walks provide a method to explore all possible paths in a parallel way, due to constructive quantum interference along the paths. Therefore, paying attention to quantum walks on hypergraphs is a natural choice.

      In this paper, we focus on discrete-time quantum walks on regular uniform hypergraphs. By analyzing the mathematical formalism of hypergraphs and three existing discrete-time quantum walks\cite{portugal2016establishing}(coined quantum walks, Szegedy's quantum walks, and staggered quantum walks), we find that discrete-time quantum walks on regular uniform hypergraphs can be transformed into Szegedy's quantum walks on bipartite graphs that are used to model the original hypergraphs. Furthermore, the mapping is one to one. That is, we can study Szegedy's quantum walks on bipartite graphs instead of the corresponding quantum walks on regular uniform hypergraphs. In Ref. \cite{szegedy2004quantum}, Szegedy proved that his schema brings about a quadratic speed-up. Hence, we construct a model for quantum walks on bipartite graphs of regular uniform hypergraphs with Szegedy's quantum walks. In the model, the evolution operator of an extended Szegedy's walks depends directly on the transition probability matrix of the Markov chain associated with the hypergraphs.

      In  more detail, we first introduce the classical random walks on hypergraphs, in order to get the vertex-edge transition matrix and the edge-vertex transition matrix. We then define a bipartite graph that is used to model the original hypergraph. Lastly, we construct quantum operators on the bipartite graph using extended Szegedy's quantum walks, which is the quantum analogue of a classical Markov chain. In this work, we deal with the case that the cardinalities of the two disjoint sets can be different from each other in the bipartite graph. In addition, we deliver a slightly different version of the spectral properties of the transition matrix, which is the essence of the quantum walks. As a result, our work generalizes quantum walks on regular uniform hypergraphs by extending the classical Markov chain, due to Szegedy's quantum walks.

    The paper is organized as follows. In Sec. II, we provide basic definitions for random walks on hypergraphs. In Sec. III, we construct a method for quantizing  Markov chain to create  discrete-time quantum walks on regular uniform hypergraphs. In Sec. IV, we analyze the eigen-decomposition of the operator. In Sec. V, we present conclusions and outlook on possible future directions.

\section{Review of random walks on hypergraphs}
\label{sec2:level1}
    We start by defining some standard definitions of a hypergraph that will be used throughout this paper. We then briefly describe random walks on hypergraphs.

\subsection{Notations}
     Let $HG=(V,E)$ denote a hypergraph, where $V$ is the vertex set of the hypergraph and $E\subset {{2}^{V}}\backslash \{\{\}\}$ is the set of hyperedges. $n=\left| V \right|$ is used to denote the number of vertices in the hypergraph and $m=\left| E \right|$ the number of hyperedges. Let $V=\{{{v}_{1}},{{v}_{2}},\cdots ,{{v}_{n}}\}$ and $E=\{{{e}_{1}},{{e}_{2}},\cdots ,{{e}_{m}}\}$. Given a hypergraph, define its incidence matrix $H\in {{R}^{n\times m}}$ as follows:
    \begin{equation}
        \begin{aligned}
                h(i,j)=\left\{ \begin{matrix}
                1 & \begin{matrix}
                if & {{v}_{i}}\in {{e}_{j}}  \\
                \end{matrix}  \\
                0 & \begin{matrix}
                if & {{v}_{i}}\notin {{e}_{j}}  \\
                \end{matrix}  \\
                \end{matrix} \right\}.
                \end{aligned}
        \label{eq:ClassificationModel}
        \end{equation}
    Note that the sum of the entries in any column is the degree of the corresponding edge. Similarly, the sum of the entries in a particular row is the degree of the corresponding vertex. Then, the vertex and hyperedge degrees are defined as follows:
    \begin{equation}
            \begin{aligned}
                d(v)=\sum\limits_{e\in E}{h(v,e)=}\left| E(v) \right|
            \end{aligned},
        \label{eq:ClassificationModel}
        \end{equation}
    \begin{equation}
            \begin{aligned}
                E(v)=\{e\in E:v\in e\}
             \end{aligned},
        \label{eq:ClassificationModel}
        \end{equation}
    \begin{equation}
            \begin{aligned}
            \delta (e)=\sum\limits_{v\in V}{h(v,e)=}\left| e \right|
            \end{aligned},
        \label{eq:ClassificationModel}
        \end{equation}
where $E(v)$ is the set of hyperedges incident to $v$. Let ${{D}_{v}}$ and ${{D}_{e}}$ denote the diagonal matrices of the degrees of the vertices and edges, respectively. A hypergraph is $d-regular$ if all its vertices have the same degree.  Also, a hypergraph is $k-uniform$ if all its hyperedges have the same cardinality. In this paper, we will restrict our reach to quantum walks on $d-regular$ and $k-uniform$ hypergraphs from now on, denoting them as $H{{G}_{k,d}}$.

\subsection{Random walks on hypergraphs}
\label{sec2.1:level2}
    A random walk on a hypergraph $HG=(V,E)$ is a Markov chain on the state space $V$ with its transition matrix $P$. The particle can move from vertex ${{v}_{i}}$ to vertex ${{v}_{j}}$ if there is a hyperedge containing both vertices. According to Ref. \cite{zhou2007learning}, a random walk on a hypergraph is seen as a two-step process. First, the particle chooses a hyperedge $e$ incident with the current vertex $v$. Then, the particle picks a destination vertex $u$ within the chosen hyperedge satisfying the following: $v,u\in e$. Therefore,the probability of moving from vertex ${{v}_{i}}$to ${{v}_{j}}$is:

    \begin{equation}
            \begin{aligned}
            {{P}_{ij}}=P({{v}_{i}},{{v}_{j}})=\sum\limits_{k=1}^{m}{\frac{{{h}_{ik}}{{h}_{jk}}}{d({{v}_{i}})\delta ({{e}_{k}})}}=\frac{1}{d({{v}_{i}})}\sum\limits_{k=1}^{m}{\frac{{{h}_{ik}}{{h}_{jk}}}{\delta ({{e}_{k}})}}
            \end{aligned},
        \label{eq:ClassificationModel}
        \end{equation}
   or, more accurately, the equation can be written as
   \begin{equation}
            \begin{aligned}
            P=\sum\limits_{\begin{smallmatrix}
    e\in E, \\
    \{v,u\}\subseteq e
    \end{smallmatrix}}{\frac{1}{d(v)\delta (e)}}=\frac{1}{d(v)}\sum\limits_{\begin{smallmatrix}
    e\in E, \\
    \{v,u\}\subseteq e
    \end{smallmatrix}}{\frac{1}{\delta (e)}}
            \end{aligned}.
        \label{eq:ClassificationModel}
        \end{equation}

   Alternately, a random walk on a hypergraph can be seen as a Markov chain on the hyperedges. At each step, the particle randomly chooses a hyperedge from the set of neighbors of the current hyperedge through the chosen vertex from the current hyperedge. Let the state space of the chain be $E$ and the transition matrix $Q$. The probability of moving form ${{e}_{i}}$ to ${{e}_{j}}$ is
   \begin{equation}
            \begin{aligned}
            {{Q}_{ij}}=Q({{e}_{i}},{{e}_{j}})=\sum\limits_{k=1}^{n}{\frac{{{h}_{ki}}{{h}_{kj}}}{\delta ({{e}_{i}})d({{v}_{k}})}}=\frac{1}{\delta ({{e}_{i}})}\sum\limits_{k=1}^{n}{\frac{{{h}_{ki}}{{h}_{kj}}}{d({{v}_{k}})}}
            \end{aligned},
        \label{eq:ClassificationModel}
        \end{equation}
or, alternatively,
   \begin{equation}
            \begin{aligned}
            Q=\sum\limits_{\begin{smallmatrix}
            v\in V, \\
            v\in e\bigcap f
            \end{smallmatrix}}{\frac{1}{\delta (e)d(v)}}=\frac{1}{\delta (e)}\sum\limits_{\begin{smallmatrix}
            v\in V, \\
            v\in e\bigcap f
           \end{smallmatrix}}{\frac{1}{d(v)}}
            \end{aligned}.
        \label{eq:ClassificationModel}
        \end{equation}

   Let ${{P}_{VE}}$ denote the vertex-edge transition matrix
    \begin{equation}
            \begin{aligned}
                {{P}_{VE}}=D_{v}^{-1}H
             \end{aligned}
        \label{eq:ClassificationModel}
        \end{equation}
   and ${{P}_{EV}}$ the edge-vertex transition matrix
      \begin{equation}
            \begin{aligned}
                {{P}_{EV}}=D_{e}^{-1}{{H}^{T}}
             \end{aligned}
        \label{eq:ClassificationModel}
        \end{equation}
   with transition probability
   \begin{equation}
            \begin{aligned}
                \sum\limits_{e\in E}{{{p}_{ve}}=1},\forall v\in V,
             \end{aligned}
        \label{eq:ClassificationModel}
        \end{equation}
   \begin{equation}
            \begin{aligned}
                \sum\limits_{v\in V}{{{p}_{ev}}=1},\forall e\in E.
             \end{aligned}
        \label{eq:ClassificationModel}
        \end{equation}
   Naturally, we can indicate $P$ and $Q$  in matrix form, respectively, as
   \begin{equation}
            \begin{aligned}
                P=D_{v}^{-1}HD_{e}^{-1}{{H}^{T}}={{P}_{VE}}{{P}_{EV}},
             \end{aligned}
        \label{eq:ClassificationModel}
        \end{equation}
   \begin{equation}
            \begin{aligned}
                Q=D_{e}^{-1}{{H}^{T}}D_{v}^{-1}H={{P}_{EV}}{{P}_{VE}}.
             \end{aligned}
        \label{eq:ClassificationModel}
        \end{equation}

\section{Quantum walks on hypergraphs}
\label{sec3:level1}
     In this section, we design quantum walks on regular uniform hypergraphs by means of Szegedy's quantum walks. We first convert the hypergraph into its associated bipartite graph, which can be used to model the hypergraph. We then define quantum operators on the bipartite graph using Szegedy's quantum walks, which are a quantization of random walks.

\subsection{Bipartite graphs model of the hypergraphs}
\label{sec3.1.1:level3}

    A hypergraph $HG$ can be represented usefully by a bipartite graph $BG$ as follows: the vertices $V$ and the edges $E$ of the hypergraph are the partitions of $BG$, and $\left( {{v}_{i}},{{e}_{j}} \right)$ are connected with an edge if and only if vertex ${{v}_{i}}$ is contained in edge ${{e}_{j}}$  in $HG$ . Formally, $B(H)=G(V\bigcup E,{{E}_{B}})$ and $\left( {{v}_{i}},{{e}_{j}} \right)\in {{E}_{B}}$ iff ${{h}_{ij}}=1$. The biadjacency matrix describing $B(H)$ is the following $\left( n+m \right)\times \left( n+m \right)$ matrix:
    \begin{equation}
            \begin{aligned}
                {{A}_{B}}=\left( \begin{matrix}0 & H  \\{{H}^{T}} & 0  \\\end{matrix} \right)
             \end{aligned},
        \label{eq:ClassificationModel}
        \end{equation}
    where $H$  with elements (1) is the incidence matrix of $HG$. Under this correspondence, the biadjacency matrices of bipartite graphs are exactly the incidence matrices of the corresponding hypergraphs. A similar reinterpretation of adjacency matrices may be used to show a one-to-one correspondence between regular uniform hypergraphs and bipartite graphs. That is, discrete-time quantum walks on regular uniform hypergraphs can be transformed into quantum walks on  bipartite graphs that are used to model the original hypergraphs.

    The transformation process is outlined in detail below. If there is a hyperedge ${{e}_{k}}$ containing both vertices ${{v}_{i}}$ and ${{v}_{j}}$ in the original hypergraph $HG=(V,E)$, convert it into two edges $({{v}_{i}},{{e}_{k}})$ and $({{e}_{k}},{{v}_{j}})$ in the bipartite graph. As a concrete example, we consider a $3-uniform$ and $2-regular$ hypergraph with the vertexes set $V=\{{{v}_{1}},{{v}_{2}},{{v}_{3}},{{v}_{4}},{{v}_{5}},{{v}_{6}}\}$ and the set of hyperedges $E=\{{{e}_{1}},{{e}_{2}},{{e}_{3}},{{e}_{4}}\}$. Then, a bipartite graph $B{{G}_{6,4}}$ with partite sets $V=\{{{v}_{1}},{{v}_{2}},{{v}_{3}},{{v}_{4}},{{v}_{5}},{{v}_{6}}\}$ and $E=\{{{e}_{1}},{{e}_{2}},{{e}_{3}},{{e}_{4}}\}$ can represent the hypergraph $H{{G}_{3,2}}$, which is depicted in Fig.1.

     \begin{figure}
       	  	\centerline{\includegraphics [width=0.65\textwidth] {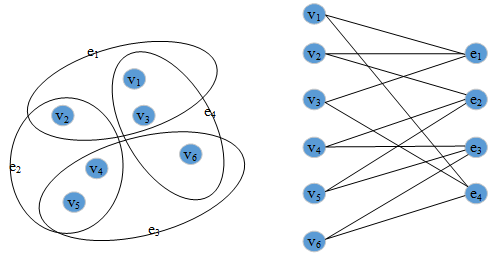}}
            \caption{Example of a hypergraph with six vertexes and four hyperedges, and its associated bipartite graph.}
         	\label{fig1:}
     	\end{figure}

    \textbf{Theorem 1 } Let $HG=(V,E)$ be a hypergraph, and we have
    \begin{equation}
            \begin{aligned}
                \sum\limits_{v\in V}{d(v)}=\sum\limits_{e\in E}{d(e)}
             \end{aligned}.
        \label{eq:ClassificationModel}
        \end{equation}
    \textbf{Proof:}  Let $B(H)=G(V\bigcup E,{{E}_{B}})$ be the incidence graph of $HG=(V,E)$. We sum the degrees in the part $E$ and in the part $V$ in $B(H)$. Since the sums of the degrees in these two parts are equal, we obtain the result.

    In particular, if the hypergraph is $d-regular$ and $k-uniform$, we obtain $nd=mk$.

\subsection{Szegedy quantum walks on the bipartite graphs}
\label{sec3.2:level2}

    Since we have transformed the hypergraph $H{{G}_{k,d}}$ into its bipartite graph $B{{G}_{n,m}}$, we now describe Szegedy quantum walks that take place on the obtained bipartite graph $B{{G}_{n,m}}$  by extending the class of possible Markov chains. The quantum walks on the hypergraph $HG$ start by considering an associated Hilbert space that is a linear subspace of the vector ${{H}^{{{n}^{2}}m}}=H_{v}^{n}\otimes H_{e}^{m}\otimes H_{v}^{n}$, where $n=\left| V \right|$, $m=\left| E \right|$. The computational basis of ${{H}^{{{n}^{2}}m}}$ is $\left\{ \left| {{v}_{i}},e,{{v}_{j}} \right\rangle :e\in E,{{v}_{i}},{{v}_{j}}\in V,{{v}_{i}},{{v}_{j}}\in e \right\}$. In addition, quantum walks on the bipartite graph $B{{G}_{n,m}}$ with biadjacent matrix (15) have an associated Hilbert space ${{H}_{A}}=H_{v}^{n}\otimes H_{e}^{m}$ and ${{H}_{B}}=H_{e}^{m}\otimes H_{v}^{n}$.

    To identify quantum analogues of Markov chains -  that is, the classical random walks with probability matrices (9) and (10) with entries of (11) and (12) - we define the vertex-edge transition operators: $A:{{H}^{n}}\to {{H}^{nd}}$ and edge-vertex transition operators: $B:{{H}^{m}}\to {{H}^{mk}}$ as follows:
    \begin{equation}
            \begin{aligned}
                A=\sum\limits_{v\in V}{\left| {{\alpha }_{v}} \right\rangle \left\langle  v \right|}
             \end{aligned},
        \label{eq:ClassificationModel}
        \end{equation}

    \begin{equation}
            \begin{aligned}
                B=\sum\limits_{e\in E}{\left| {{\beta }_{e}} \right\rangle \left\langle  e \right|}
             \end{aligned},
        \label{eq:ClassificationModel}
        \end{equation}
where
     \begin{equation}
            \begin{aligned}
                \left| {{\alpha }_{v}} \right\rangle =\left| v \right\rangle \otimes \left( \sum\limits_{e\in E}{\sqrt{{{p}_{ve}}}\left| e \right\rangle } \right)
             \end{aligned},
        \label{eq:ClassificationModel}
        \end{equation}

     \begin{equation}
            \begin{aligned}
                \left| {{\beta }_{e}} \right\rangle =\left( \sum\limits_{v\in V}{\sqrt{{{p}_{ev}}}\left| e \right\rangle } \right)\otimes \left| v \right\rangle
             \end{aligned}.
        \label{eq:ClassificationModel}
        \end{equation}
     The transition operators are defined on the Hilbert space ${{H}_{A}}$ and ${{H}_{B}}$ separately, where the computational basis of ${{H}^{nd}}$ is $\left\{ \left| v,e \right\rangle :v\in V,e\in E \right\}$ and the computational basis of ${{H}^{mk}}$ is $\left\{ \left| e,v \right\rangle :e\in E,v\in V \right\}$. The states $\left| {{\alpha }_{v}} \right\rangle$ and $\left| {{\beta }_{e}} \right\rangle$ as superpositions that start from vertex $v$ to hyperedge $e$ and from hyperedge $e$ to vertex $v$, respectively. Obviously, the dimensions of $A$ and $B$ are $nd\times n$ and $mk\times m$, respectively. Note that $nd=mk$ from theorem 1. Using (19) and (20) along with (11) and (12), we obtain the following properties:
     \begin{equation}
            \begin{aligned}
               \langle  {{\alpha }_{v}} \left| {{\alpha }_{v'}} \right\rangle ={{\delta }_{vv'}}
             \end{aligned},
        \label{eq:ClassificationModel}
        \end{equation}
     \begin{equation}
            \begin{aligned}
                \langle  {{\beta }_{e}} \left| {{\beta }_{e'}} \right\rangle ={{\delta }_{ee'}}
             \end{aligned},
        \label{eq:ClassificationModel}
        \end{equation}
as well as
     \begin{equation}
          \begin{aligned}
          {{A}^{T}}A ={{I}_{n}}
          \end{aligned},
          \label{}
     \end{equation}
     \begin{equation}
          \begin{aligned}
          {{B}^{T}}B ={{I}_{m}}
          \end{aligned}.
          \label{}
     \end{equation}
     One can easily verify that $\left| {{\alpha }_{v}} \right\rangle$  and $\left| {{\beta }_{e}} \right\rangle$ are unit vectors due to the stochasticity of ${{P}_{VE}}$ and ${{P}_{EV}}$. Distinctly, these equations imply that the action of $A$ preserves the norm of the vectors. The same is true regarding $B$.

     We now immediately define the projectors ${{\Pi }_{A}}$ and ${{\Pi }_{B}}$ as follows:

     \begin{equation}
            \begin{aligned}
                {{\Pi }_{A}}=A{{A}^{T}}=\sum\limits_{v\in V}{\left| {{\alpha }_{v}} \right\rangle \left\langle  {{\alpha }_{v}} \right|}
             \end{aligned},
        \label{eq:ClassificationModel}
        \end{equation}

     \begin{equation}
            \begin{aligned}
                {{\Pi }_{B}}=B{{B}^{T}}=\sum\limits_{e\in E}{\left| {{\beta }_{e}} \right\rangle \left\langle  {{\beta }_{e}} \right|}
             \end{aligned}.
        \label{eq:ClassificationModel}
        \end{equation}
     Using Eqs.(25) and (26), it is easy to see that ${{\Pi }_{A}}$ projects onto subspace ${{H}_{A}}$ spanned by $\left\{ \left| {{\alpha }_{v}} \right\rangle :v\in V \right\}$, and ${{\Pi }_{B}}$ projects onto subspace ${{H}_{B}}$ spanned by $\left\{ \left| {{\beta }_{e}} \right\rangle :e\in E \right\}$.
     After obtaining the projectors, we can define the associated reflection operators, which are
     \begin{equation}
            \begin{aligned}
                {{R}_{A}}=2{{\Pi }_{A}}-{{I}_{nd}}
             \end{aligned},
        \label{eq:ClassificationModel}
        \end{equation}
     \begin{equation}
            \begin{aligned}
                {{R}_{B}}=2{{\Pi }_{B}}-{{I}_{mk}}
             \end{aligned},
        \label{eq:ClassificationModel}
        \end{equation}
where ${{I}_{nd}}={{I}_{mk}}$ is the identity operator.  ${{R}_{A}}$ is the reflection though the line generated by $\left| {{\alpha }_{v}} \right\rangle$, and ${{R}_{B}}$ is the reflection though the line generated by $\left| {{\beta }_{e}} \right\rangle$. Note that the reflection operators ${{R}_{A}}$ and ${{R}_{B}}$ are unitary and Hermitian. With all the information, a single step of the quantum walks is given by the unitary evolution operator
     \begin{equation}
            \begin{aligned}
                W={{R}_{B}}{{R}_{A}}
             \end{aligned}
        \label{eq:ClassificationModel}
        \end{equation}
based on the transition matrix $P$. In the bipartite graph, an application of $W$ corresponds to two quantum steps of the walk from $v$ to $e$ and from $e$ to $v$. At time $t$, the whole operator of the quantum walks is ${{W}^{t}}$.

\section{Spectral analysis of quantum walks on hypergraphs}

    In many classical algorithms, the eigen-spectrum of the transition matrix $P$ plays a critical role in the analysis of Markov chains. In a similar way, we now proceed to study the quantitative spectrum of the quantum walks unitary operator $W$.

    Szegedy proved a spectral theorem for quantum walks, $W=re{{f}_{2}}re{{f}_{1}}$, in Ref. \cite{szegedy2004quantum}. In this section, we deliver a slightly different version in that the cardinality of set $X$ may be different from the cardinality of set $Y$ in the bipartite graph. In order to analyze the spectrum, we need to study the spectral properties of an $n\times m$  matrix $D$, which indeed establishes a relation between the classical Markov chains and the quantum walks. This matrix is defined as follows:

    \textbf{(Discriminant Matrix) } The discriminant matrix for $W$ is
    \begin{equation}
            \begin{aligned}
                {{D}_{nm }}={{A}^{T}}B
             \end{aligned}.
        \label{eq:ClassificationModel}
        \end{equation}
    Herein, suppose that $n\ge m$. Also, it follows from the definition that $D=\sqrt{{{P}_{VE}}\circ {{P}_{EV}}}$ with entries
    \begin{equation}
            \begin{aligned}
                {{D}_{ve}}=\sqrt{{{p}_{ve}}{{p}_{ev}}},\forall v\in V,\forall e\in E
             \end{aligned}.
        \label{eq:ClassificationModel}
        \end{equation}

    Suppose that the discriminant matrix $D$ has the singular value decomposition $D=U\Sigma {{V}^{T}}=\sum\nolimits_{i}{{{\sigma }_{i}}}{{\mu }_{i}}\nu _{i}^{T}$. The left singular vectors $\left| {{\mu }_{k}} \right\rangle $ satisfy
    \begin{equation}
            \begin{aligned}
                D\left| {{\nu }_{k}} \right\rangle ={{\sigma }_{k}}\left| {{\mu }_{k}} \right\rangle
             \end{aligned}
        \label{eq:ClassificationModel}
        \end{equation}
    and the right singular vectors
    \begin{equation}
            \begin{aligned}
                \left\langle  {{\mu }_{k}} \right|D=\left\langle  {{\nu }_{k}} \right|{{\sigma }_{k}}
             \end{aligned}
        \label{eq:ClassificationModel}
        \end{equation}
    with ${{\sigma }_{k}}$ the singular value.

     \textbf{Theorem 2 } For any ${{\sigma }_{k}}$ the singular value of $D$, $0\le {{\sigma }_{k}}\le 1$.

    \textbf{Proof: } First, let $Dk={{\sigma }_{k}}k$.
    Then we obtain
    \begin{equation}
        \begin{aligned}
           {{\left| {{\sigma }_{k}} \right|}^{2}}{{\left\| k \right\|}^{2}}& ={{\left\| Dk \right\|}^{2}} \\
           & =\left\langle Dk,Dk \right\rangle  \\
           & =\left\langle {{A}^{T}}Bk,{{A}^{T}}Bk \right\rangle  \\
           & =\left\langle Bk,A{{A}^{T}}Bk \right\rangle  \\
           & \le \left\langle Bk,Bk \right\rangle  \\
           & =\left\langle k,{{B}^{T}}Bk \right\rangle  \\
           & =\left\langle k,k \right\rangle ={{\left\| k \right\|}^{2}}.
        \end{aligned}
        \label{}
    \end{equation}
    Thus, $\left| {{\sigma }_{k}} \right|\le 1$. Since $\left\langle k,{{D}^{T}}Dk \right\rangle \ge 0$ for all $k$, we have $0\le {{\sigma }_{k}}$. Therefore, $0\le {{\sigma }_{k}}\le 1$.

    Observing theorem 2 , we can write the singular value ${{\sigma }_{k}}$ as $\cos {{\theta }_{k}}$, where ${{\theta }_{k}}$ is the principal angle between subspace ${{H}_{A}}$ and ${{H}_{B}}$. In the early literature \cite{bjorck1973numerical}, Bj{\"o}rck and Golub deducted the relationship between the singular value decomposition and the principal angle ${{\theta }_{k}}$ between subspace ${{H}_{A}}$ and ${{H}_{B}}$. That is, $\cos ({{\theta }_{k}})={{\sigma }_{k}}$.

    In the remainder of this section, we will explore the eigen-decomposition of the operator $W$,  which can be calculated from the singular value decomposition of $D$.

    Using $A$ to left-multiply (32) and $B$ to left-multiply (33), We have
    \begin{equation}
            \begin{aligned}
                AD\left| {{\nu }_{k}} \right\rangle = \left( A{{A}^{T}} \right)B\left| {{\nu }_{k}} \right\rangle ={{\sigma }_{k}}A\left| {{\mu }_{k}} \right\rangle
             \end{aligned},
        \label{eq:ClassificationModel}
        \end{equation}
    \begin{equation}
            \begin{aligned}
                B{{D}^{T}}\left| {{\mu }_{k}} \right\rangle =\left( B{{B}^{T}} \right)A\left| {{\mu }_{k}} \right\rangle ={{\sigma }_{k}}B\left| {{\nu }_{k}} \right\rangle
             \end{aligned}.
        \label{eq:ClassificationModel}
        \end{equation}
    As we know, the action of $A$ and $B$ preserve the norm of the vectors, and $\left| {{\nu }_{k}} \right\rangle $ and $\left| {{\mu }_{k}} \right\rangle$ are unit vectors, so $A\left| {{\mu }_{k}} \right\rangle$ and $B\left| {{\nu }_{k}} \right\rangle$ also are unit vectors. Further, (35) and (36) imply that ${{\Pi }_{A}}$ and ${{\Pi }_{B}}$ have a symmetric action on $A\left| {{\mu }_{k}} \right\rangle$ and $B\left| {{\nu }_{k}} \right\rangle$. Therefore, we can conclude that the subspace $span\{A\left| {{\mu }_{k}} \right\rangle ,B\left| {{\nu }_{k}} \right\rangle \}$ is invariant under the action of ${{\Pi }_{A}}$ and ${{\Pi }_{B}}$.

    We then have
    \begin{equation}
    \begin{aligned}
         WA\left| {{\mu }_{k}} \right\rangle & ={{R}_{B}}{{R}_{A}}A\left| {{\mu }_{k}} \right\rangle  \\
         & ={{R}_{B}}A\left| {{\mu }_{k}} \right\rangle  \\
         & =2B{{B}^{T}}A\left| {{\mu }_{k}} \right\rangle -A\left| {{\mu }_{k}} \right\rangle  \\
         & =2{{\sigma }_{k}}B\left| {{\nu }_{k}} \right\rangle -A\left| {{\mu }_{k}} \right\rangle
    \end{aligned}
    \label{}
    \end{equation}
    and
    \begin{equation}
    \begin{aligned}
          WB\left| {{\nu }_{k}} \right\rangle& ={{R}_{B}}{{R}_{A}}B\left| {{\nu }_{k}} \right\rangle  \\
          & ={{R}_{B}}(2A{{A}^{T}}-I)B\left| {{\nu }_{k}} \right\rangle  \\
          & ={{R}_{B}}(2A{{A}^{T}}B\left| {{\nu }_{k}} \right\rangle )-B\left| {{\nu }_{k}} \right\rangle  \\
          & ={{R}_{B}}(2{{\sigma }_{k}}A\left| {{\mu }_{k}} \right\rangle )-B\left| {{\nu }_{k}} \right\rangle  \\
          & =2{{\sigma }_{k}}(2{{\sigma }_{k}}B\left| {{\nu }_{k}} \right\rangle -A\left| {{\mu }_{k}} \right\rangle )-B\left| {{\nu }_{k}} \right\rangle  \\
          & =\left( 4\sigma _{k}^{2}-1 \right)B\left| {{\nu }_{k}} \right\rangle -2{{\sigma }_{k}}A\left| {{\mu }_{k}} \right\rangle.
    \end{aligned}
    \label{}
    \end{equation}
    Hence, we can conclude that the subspace $span\{A\left| {{\mu }_{k}} \right\rangle ,B\left| {{\nu }_{k}} \right\rangle \}$ is invariant under the action $W$. This, in turn, helps us characterize the eigenvalues of $W$ using the singular values of $D$.

    Suppose that
    \begin{equation}
    \begin{aligned}
         W\left| k \right\rangle ={{\lambda }_{k}}\left| k \right\rangle
    \end{aligned}
    \label{}
    \end{equation}
    and
    \begin{equation}
    \begin{aligned}
         \left| k \right\rangle =aA\left| {{\mu }_{k}} \right\rangle +bB\left| {{\nu }_{k}} \right\rangle
    \end{aligned}.
    \label{}
    \end{equation}
    Simply plugging (40) into formulas (39), we obtain the following equation:

    \begin{equation}
    \begin{aligned}
         W\left| k \right\rangle ={{\lambda }_{k}}aA\left| {{\mu }_{k}} \right\rangle +{{\lambda }_{k}}bB\left| {{\nu }_{k}} \right\rangle
    \end{aligned}.
    \label{}
    \end{equation}
    Then, left-multiplying (40) by $W$, we have
    \begin{equation}
    \begin{aligned}
         W\left| k \right\rangle & =W(aA\left| {{\mu }_{k}} \right\rangle +bB\left| {{\nu }_{k}} \right\rangle ) \\
         & =aWA\left| {{\mu }_{k}} \right\rangle +bWB\left| {{\nu }_{k}} \right\rangle  \\
         & =-(a+2b{{\sigma }_{k}})A\left| {{\mu }_{k}} \right\rangle +[2a\sigma _{k}^{2}+b(4{{\sigma }_{k}}-1)]B\left| {{\nu }_{k}} \right\rangle.
    \end{aligned}
    \label{}
    \end{equation}
    Comparing formulas (41) and (42), we can obtain the following  equations:
    \begin{equation}
    \begin{aligned}
         {{\lambda }_{k}}a=-(a+2b{{\sigma }_{k}})
    \end{aligned},
    \label{}
    \end{equation}
    \begin{equation}
    \begin{aligned}
         {{\lambda }_{k}}b=2a\sigma _{k}^{2}+b(4{{\sigma }_{k}}-1)
    \end{aligned}.
    \label{}
    \end{equation}

     Concerning  unit vectors $A\left| {{\mu }_{k}} \right\rangle$ and $B\left| {{\nu }_{k}} \right\rangle$, we consider two cases with respect to non-collinearity and collinearity, as follows.

     Case 1. First, we consider  that $A\left| {{\mu }_{k}} \right\rangle$ and $B\left| {{\nu }_{k}} \right\rangle$ are linearly independent.

     Using ${{\sigma }_{k}}=\cos {{\theta }_{k}}$, we obtain
     \begin{equation}
    \begin{aligned}
         {{\lambda }_{k}}={{e}^{\pm 2i{{\theta }_{k}}}}
    \end{aligned}
    \label{}
    \end{equation}
   through a series of algebraic operations. Furthermore, we have the corresponding eigenvectors
    \begin{equation}
    \begin{aligned}
         \left| k \right\rangle =\frac{A\left| {{\mu }_{k}} \right\rangle -{{e}^{\pm i{{\theta }_{k}}}}B\left| {{\nu }_{k}} \right\rangle }{\sqrt{2}\sin {{\theta }_{k}}}
    \end{aligned}.
    \label{}
    \end{equation}

    Case 2. Then, we consider  that $A\left| {{\mu }_{k}} \right\rangle$ and $B\left| {{\nu }_{k}} \right\rangle$ are collinear. However, since $A\left| {{\mu }_{k}} \right\rangle$ is invariant under the action of ${{\Pi }_{A}}$, $B\left| {{\nu }_{k}} \right\rangle$ also is; and vice versa, since $B\left| {{\nu }_{k}} \right\rangle$ is invariant under ${{\Pi }_{B}}$, and $A\left| {{\mu }_{k}} \right\rangle $ also is. Therefore, $A\left| {{\mu }_{k}} \right\rangle$ and $B\left| {{\nu }_{k}} \right\rangle$ are invariant under the action of $W$, and $A\left| {{\mu }_{k}} \right\rangle$ are eigenvectors of $W$ with eigenvalue 1.

    Now, we turn to the dimensionality of the spaces. We learned earlier that $nd$ is the dimension of edge Hilbert space about the bipartite graph $B{{G}_{n,m}}$, and the discriminant matrix $D$ has $m$ singular values, only some of which are non-zero. Space ${{H}_{A}}$ spanned by $\left\{ \left| {{\alpha }_{v}} \right\rangle :v\in V \right\}$, and space ${{H}_{B}}$ spanned by $\left\{ \left| {{\beta }_{e}} \right\rangle :e\in E \right\}$, are $n-dimension$ and $m-dimension$ subspaces of ${{H}^{nd}}$, respectively. Let ${{H}_{AB}}$ be the space spanned by $\left\{ \left| {{\alpha }_{v}} \right\rangle :v\in V \right\}$ and $\left\{ \left| {{\beta }_{e}} \right\rangle :e\in E \right\}$. Then, the dimension of ${{H}_{AB}}$ is $m+n$, when $A\left| {{\mu }_{k}} \right\rangle$ and $B\left| {{\nu }_{k}} \right\rangle$ are linearly independent. On the other hand, the dimension of ${{H}_{AB}}$ is $n-m$. Therefore, the operator $W$ has $nd-(m+n)$ eigenvalues 1 and $n-m$ eigenvalues -1 in the one-dimensional subspaces invariant, and  $2m$ eigenvalues in the two-dimensional subspaces. Table 1 gives the eigenvalues of $W$ and the singular values of $D$ up to now.

   \begin{table}[tbp]
     \centering
     \caption{Eigenvalues of $W$ obtained from the singular values of $D$, and angles ${{\theta }_{k}}$ obtained from the formula ${{\sigma }_{k}}=\cos {{\theta }_{k}}$, where $k=1,2,\cdots ,m$.}
     \begin{tabular}{|c|c|c|}
        \hline
        Number of the eigenvalue &Eigenvalue of $W$ &Singular values of $D$ \\
        \hline
        $2m$   &${{\lambda }_{k}}={{e}^{\pm 2i{{\theta }_{k}}}}$                                    &${{\sigma }_{k}}=\cos {{\theta }_{k}}$ \\
                 $nd-(m+n)$  &1  &1 \\
                 $n-m$  &0  &-1\\
        \hline
    \end{tabular}
    \end{table}

    As a consequence, we obtain the following theorem:

    \textbf{Theorem 3 }  Let $W$ be the unitary evolution operator on $B{{G}_{n,m}}$ . Suppose that $n\ge m$. Then $W$ has $nd-(m+n)$ eigenvalues 1 and $n-m$ eigenvalues -1 in the one-dimensional invariant subspaces, and $2m$ eigenvalues in the two-dimensional subspaces. The $2m$ eigenvalues are ${{\lambda }_{k}}={{e}^{\pm 2i{{\theta }_{k}}}}$($0<{{\theta }_{k}}<\frac{\pi }{2}$) where ($k=1,2,\cdots ,m$ ) and the eigenvectors $\left| k \right\rangle =\frac{A\left| {{\mu }_{k}} \right\rangle -{{e}^{\pm i{{\theta }_{k}}}}B\left| {{\nu }_{k}} \right\rangle }{\sqrt{2}\sin {{\theta }_{k}}}$.

\section{Conclusions and outlook}
\label{sec7:level1}

    Quantum walks are one of the elementary techniques of developing quantum algorithms. The development of successful quantum walks on graphs-based algorithms have boosted such areas as element distinctness, searching for a marked element, and graph isomorphism. In addition, the utility of walking on hypergraphs has been probed deeply in several contexts, including natural language parsing, social networks database design, or image segmentation, and so on. Therefore, we put our attention on quantum walks on hypergraphs considering its promising power of inherent parallel computation.

    In this paper, we developed a new schema for discrete-time quantum walks on regular uniform hypergraphs using extended Szegedy's walks that naturally quantize classical random walks and yield quadratic speed-up compared to the hitting time of classical random walks. We found the one-to-one correspondence between regular uniform hypergraphs and bipartite graphs. Through the correspondence, we convert the regular uniform hypergraph into its associated bipartite graph on which extended Szegedy's walks take place. In addition, we dealt with the case that the cardinality of the two disjoint sets may be different from each other in the bipartite graphs. Furthermore, we delivered spectral properties of the transition matrix, which is the essence of quantum walks, and which has prepared for followup studies.

    Our work presents a model for quantum walks on regular uniform hypergraphs, and the model opens the door to quantum walks on hypergraphs. We hope our model can inspire more fruitful results in quantum walks on hypergraphs. Our model provides the foundation for building up quantum algorithms on the strength of quantum walks on hypergraphs. Moreover, the algorithms of quantum walks on hypergraphs will be useful in quantum computation such as quantum walks search, quantized Google's PageRank, and quantum machine learning, based on hypergraphs.

    Based on the preliminary research presented here, the following areas require further investigation:

    1  Since the quantum walk evolutions are unitary, the probability of finding an element will oscillate through time. Therefore, the hitting time must be close to the time where the probability seems to peak for the very first time. One can calculate analytically the hitting time and the probability of finding a set of marked vertices on the hypergraphs using Szegedy' s quantum hitting time.

    2  We have defined discrete-time quantum walks on regular uniform hypergraphs, while the condition can be relaxed. If the hypergraph is any one of several types, the questions then become a) how to generalize the quantum walks from regular uniform hypergraphs to any hypergraphs, and b) how to define quantum walks on hypergraphs?

\begin{acknowledgments}
    I would like to thank Juan Xu , Yuan Su and Iwao Sato for helpful discussions. This work was supported by the Funding of National Natural Science Foundation of China (Grant No. 61571226,61701229), and the Natural Science Foundation of Jiangsu Province, China (Grant No. BK20170802).
\end{acknowledgments}


\end{document}